\documentclass[aps,prd,preprint,superscriptaddress,tightenlines,nofootinbib]{revtex4}

\usepackage{graphicx}% Include figure files
\usepackage{dcolumn}% Align table columns on decimal point
\usepackage{bm}% bold math

\begin{document}
%\preprint line(s) will be ignored for PRL/PRD
\preprint{CLNS 06/1985}       % for CLNS notes
\preprint{CLEO 06-25}         % for CLNS notes

\title{Search for invisible decays of the {\boldmath $\Upsilon(1S)$} resonance}

\author{P.~Rubin}
\affiliation{George Mason University, Fairfax, Virginia 22030}
\author{C.~Cawlfield}
\author{B.~I.~Eisenstein}
\author{I.~Karliner}
\author{D.~Kim}
\author{N.~Lowrey}
\author{P.~Naik}
\author{M.~Selen}
\author{E.~J.~White}
\author{J.~Wiss}
\affiliation{University of Illinois, Urbana-Champaign, Illinois 61801}
% ...
%\author{(CLEO Collaboration)} %FOR PRD_SPECIAL_CHANGEME
\collaboration{CLEO Collaboration} %FOR PRL,CLNS, PRDRC
\noaffiliation

\date{December 21, 2006}

\begin{abstract} 
We present a measurement of the branching fraction of invisible $\Upsilon(1S)$ decays, using 1.2 fb$^{-1}$ of data collected at the $\Upsilon(2S)$ resonance with the CLEO III detector at CESR. After subtracting  expected backgrounds from events that pass selection criteria for invisible $\Upsilon(1S)$ decay in $\Upsilon(2S) \to \pi^+ \pi^- \Upsilon(1S)$, we deduce a $90\%$ C.L. upper limit of ${\cal B}[\Upsilon(1S) \to {\rm invisible}] < 0.39 \%$.
\end{abstract}

\maketitle

Invisible decays of quarkonia to final state particles that cannot be detected in general purpose particle detectors provide a window on physics beyond the Standard Model. This is because the only allowed invisible decay of quarkonium states in the Standard Model is the decay to $\nu\bar{\nu}$ via annihilation into a virtual $Z^0$ boson. The predicted branching fraction for $\Upsilon(1S)$ is
\begin{equation}
{\cal B}[\Upsilon(1S) \to \nu\bar{\nu}] = 4.14 \times 10^{-4} \times {\cal B}[\Upsilon(1S) \to e^+e^-]
\end{equation}
or of order $10^{-5}$. Although this decay mode itself is sensitive to new physics such as the presence of an extra $Z'$ gauge boson or R-parity violating effects in supersymmetric theories, the branching fraction remains well below current experimental sensitivity \cite{nunubar}. 

On the other hand, most dark matter scenarios require some coupling between the Standard Model sector and dark matter, which could significantly increase the invisible decay rate of heavy quarkonia \cite{McElrath,Fayet}. Although the leading dark matter candidate in different theoretical models (e.g., in the Minimal Supersymmetric Standard Model) is the lightest supersymmetric particle with a mass typically $M_{\chi} > 6$ GeV/$c^2$, there is some evidence that the dark matter constituents of the universe may be as light as 100 MeV/$c^2$ and several models can be constructed to accommodate such particles. 
Based on a model-independent calculation, using only the result from the Wilkinson Microwave Anisotropy Probe (WMAP) \cite{WMAP} on the relic density of the universe, the author of \cite{McElrath} has predicted the decay branching fraction of the $\Upsilon(1S)$ to a pair of dark matter particles to be
\begin{equation}
{\cal B}[\Upsilon(1S) \to \chi\chi] \approx 0.41\%~.
\end{equation}
Kinematic factors arising from the mass of the dark matter particles or the mediator can either enhance or suppress this branching fraction. Invisible $\Upsilon$ decays can produce a pair of dark matter particles with mass less than $M_{\Upsilon}/2$ assuming the decay is mediated by a vector boson.

Upper limits have been set on the invisible decays of the $\eta$ and  $\eta^{\prime}$ from $J/\psi \to \phi\eta^{(\prime)}$ decay by the BES Collaboration \cite{BES}. In addition, a 90\% confidence level upper limit of 0.25\% on the invisible branching fraction of $\Upsilon(1S)$ from $\Upsilon(3S) \to \pi^+\pi^-\Upsilon(1S)$ decay has been reported quite recently by the Belle Collaboration \cite{BELLE}, which is an order of magnitude better than earlier limits set by the CLEO and ARGUS Collaborations \cite{oldY1S}.

In order to observe the invisible decays of the $\Upsilon(1S)$ resonance, we use the $\Upsilon(2S) \to \pi^+\pi^- \Upsilon(1S)$ transition and infer the presence of the $\Upsilon(1S)$ resonance from the missing mass in the event. In this analysis, we use 1.2 fb$^{-1}$ data collected at the peak of the $\Upsilon(2S)$ resonance with the CLEO III detector operating at the Cornell Electron Storage Ring, a symmetric $e^+e^-$ collider. The data represent about 8.7 million $\Upsilon(2S)$ resonance decays, 18.8\% of them decaying to $\pi^+\pi^-\Upsilon(1S)$ \cite{PDG}.

The CLEO III detector has excellent charged particle tracking, electromagnetic calorimetry, and particle identification capabilities, with solid angle coverage for charged and neutral particles in the polar angle range $|\cos \theta| < 0.93$ (the polar angle is measured with respect to the beam axis). The tracking system consists of a four-layer double-sided silicon vertex detector and a 47-layer drift chamber \cite{DR3} residing in a superconducting solenoid that produces a 1.5 T axial field. The CsI crystal calorimeter and the muon detection system are the same used in the CLEO II detector \cite{CLEO2}. 
Measurement of specific ionization ($dE/dx$) in the drift chamber provides particle identification.  A  Ring Imaging Cherenkov detector \cite{RICH} outside the drift chamber complements the $dE/dx$ measurements, but is not effective at the particle momenta of this analysis.

The CLEO III trigger \cite{Trigger} makes decisions based on AXIAL trigger tracks (observed in the inner 16 axial drift chamber layers), STEREO trigger tracks (which extend into the outer 31 stereo drift chamber layers), and cluster information from the calorimeter. 

When the $\Upsilon(1S)$ resonance decay is invisible, only the two soft pions from the $\Upsilon(2S)$ transition can trigger the data acquisition system. The only CLEO III trigger requirement that is sensitive to this signal is the two-track trigger, which only demands the presence of at least two AXIAL trigger tracks. 
However, the two-track trigger has been prescaled by a factor of 20 in this data set, i.e., events that satisfy the two-track trigger requirement are counted by a prescale counter and the two-track trigger bit is only set in every 20th such event. This prescaling was implemented during all data taking at the Upsilon resonances in order to reduce the trigger rate on beam-related backgrounds.

In order to study our selection efficiency for signal and possible background contributions, we use Monte Carlo (MC) simulations. The signal MC sample is created with the EvtGen \cite{EvtGen} generator using the decay sequence $\Upsilon(2S) \to \pi^+\pi^-\Upsilon(1S)$ followed by $\Upsilon(1S) \to \nu_e\bar{\nu}_e$. The di-pion transition is generated assuming relative $S$-wave between the $\pi\pi$ system and the $\Upsilon(1S)$ as well as between the two pions. Final state radiation is simulated with PHOTOS \cite{FSR}. Decays of $\Upsilon(2S)$ and continuum ($e^+e^- \to q\bar{q}, q=u,d,s,c$) events are generated with a  customized version of the JETSET \cite{JETSET} program, while non-resonant tau-pairs ($e^+e^- \to \tau^+\tau^-$) are generated using the Koralb/Tauola \cite{Koralb,Tauola} event generator. The generated events pass through a GEANT-based \cite{GEANT} full detector simulation and the same reconstruction procedure used for the data.
                                        
The analysis strategy is straightforward. We first select an inclusive sample of $\Upsilon(2S) \to \pi^+\pi^-\Upsilon(1S)$ events by finding the pion candidates. This sample represents all possible $\Upsilon(1S)$ decays, including invisible decays. Then, we select the exclusive subsample of these for which the $\Upsilon(1S)$ decays to undetected particles. 
Events with an $\Upsilon(1S)$ can be identified by the recoil mass against the di-pion (equivalent to the missing mass if only the two pions are considered in the event), which produces a peak at the $\Upsilon(1S)$ mass above a smooth combinatoric background. The total number of $\pi^+\pi^- \Upsilon(1S)$ events detected in the inclusive sample is extracted using a fit to the $\pi\pi$ recoil mass spectrum. A similar exercise is performed upon the subset of the inclusive sample with no visible particles aside from the two cascade pions. The branching fraction of invisible decays is then determined as the ratio of these two yields after background subtraction.

In order to select the inclusive sample of $\pi^+\pi^-\Upsilon(1S)$ events, we require the two pion candidates to be oppositely charged tracks with good quality fit to a helix and with polar angle $|\cos \theta| < 0.93$. The tracks must originate from the interaction point within 5 mm in the plane perpendicular to the beam axis and 5 cm along the beam axis (luminous volume). The two-track trigger bit must be set in the event. In order to ensure that each of the two pions produce an AXIAL trigger track with high probability and together satisfy the two-track trigger requirement, we demand that the transverse momentum of each pion candidate is more than 150 MeV/$c$. We also demand that the magnitude of the momentum is less than 600 MeV/$c$.
The recoil mass against the di-pion candidate is calculated using the 4-momentum of the incident beams in the lab frame ($p_{cm}^{\mu} = p_{e^-}^{\mu} + p_{e^+}^{\mu}$), and constructing the 4-product
\begin{equation}
(M_{\rm rec})^2 = (p_{cm} - p_{\pi\pi})^{\mu} (p_{cm} - p_{\pi\pi})_{\mu} ~~;
\end{equation}
$M_{\rm rec}$ is required to be within 30 MeV/$c^2$ of the nominal $\Upsilon(1S)$ mass ($9.460$ GeV/$c^2$).

A kinematic fit constraining the pion candidates to have a common vertex is performed.  We require this vertex to have a fit of reasonable quality and be in the luminous volume previously described. The improved momenta of the pions from the vertex fit are used to re-calculate $M_{\rm rec}$, slightly improving the resolution in this variable.

The dominant combinatoric background in the exclusive sample with invisible $\Upsilon(1S)$ decay arises from two-photon fusion events ($e^+ e^- \to e^+ e^- \gamma^{\star}\gamma^{\star}$) when the photons radiated by the incoming electron/positron beams produce a lepton pair or charged pion pair: $\gamma^{\star}\gamma^{\star} \to e^+e^-/\mu^+\mu^-/\pi^+\pi^-$. Because the photon emission probability peaks in the forward direction, the electron/positron beam particles tend to scatter at a very small angle and leave the interaction region undetected.
Thus only the lepton or pion pair produced by the two-photon fusion is detected (untagged two-photon events). The two-photon system is typically boosted along the beam axis and therefore the total momentum of the particles produced by the photon fusion tends to be parallel with the beam axis (i.e., the total transverse momentum is close to zero).

In order to reduce the background due to untagged two-photon fusion events, we apply a cut on the polar angle of the di-pion momentum ($|\cos\theta_{\pi\pi}| < 0.9$). In addition, we apply particle identification on the pion candidates using the specific-ionization ($dE/dx$) in the drift chamber by requiring a three standard deviation consistency with the pion hypothesis. This latter requirement is especially effective at suppressing the remaining two-photon fusion events which produce electron pairs.

We accept multiple di-pion candidates per event if they pass the di-pion selection criteria. Random combinations of two oppositely charged tracks other than the real di-pion from the $\Upsilon(2S) \to \pi^+\pi^- \Upsilon(1S)$ transition can only contribute to the smooth, slowly varying background underneath the peak; they are subtracted when we extract the number of $\pi^+\pi^-\Upsilon(1S)$ events from the fit to the recoil mass distribution.

\begin{figure}
	\centering
	\includegraphics[width=5.5in]{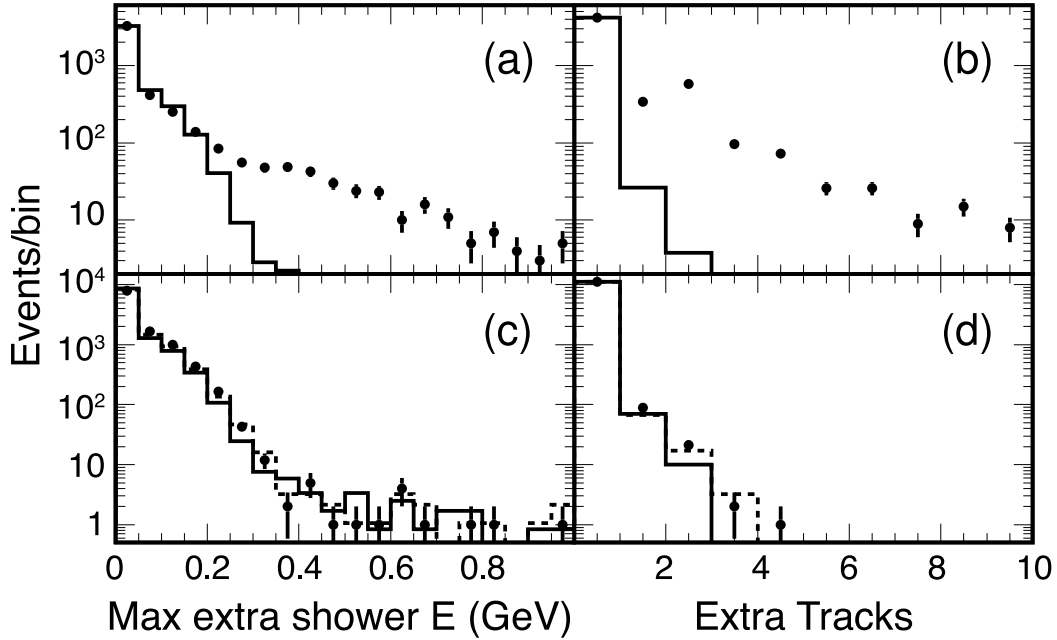}
	\caption
{The energy of the largest extra shower (left) and the number of extra tracks (right) in data and simulation. An event can enter in one of these plots only if it passes the di-pion selection and the cut on the other variable. Plots (a) and (b) compare the {\it inclusive} data (points) to {\it signal} MC simulation (solid histogram), normalized to the first data bin. Plots (c) and (d) compare {\it fully reconstructed} $\Upsilon(2S) \to \pi^+\pi^- \Upsilon(1S)$, $\Upsilon(1S) \to e^+ e^-/\mu^+\mu^-$ data and the MC simulations for such di-lepton final states (dashed histogram) and the signal process (solid histogram), both normalized to the data area.}
	\label{fig:Fig1}
\end{figure}

The exclusive sample of invisible $\Upsilon(1S)$ decays is selected from the initial inclusive $\pi^+\pi^-\Upsilon(1S)$ sample by the requirement that there is no additional good track coming from the vicinity of the interaction region and no extra good shower with more than $250$ MeV energy in the event. Tracks with similar quality criteria to the pion candidates but with momentum 50 $<p<$ 5500 MeV/$c$ are counted as "good" tracks. Showers are considered "good" if they are not associated with the pion candidates.
Figure \ref{fig:Fig1} displays, for both data and MC simulations, the distribution of the maximum extra shower energy and the number of extra tracks for events that passed the di-pion selection and the requirement on the other variable.

Based on the signal MC simulation, the efficiency of the exclusive selections on invisible $\Upsilon(1S)$ decays that passed the inclusive di-pion selections without trigger requirement is $98.8$\%. This efficiency is in good agreement with that obtained from fully reconstructed $\Upsilon(2S) \to \pi^+\pi^- \Upsilon(1S)$, $ \Upsilon(1S) \to e^+ e^-/\mu^+\mu^-$ data and MC simulation, as demonstrated in Fig.~\ref{fig:Fig1}c-d. In such fully reconstructed events, only tracks and showers not associated with any of the four final state particles are counted as extra.

\begin{figure}
	\centering
	\includegraphics[width=4.5in]{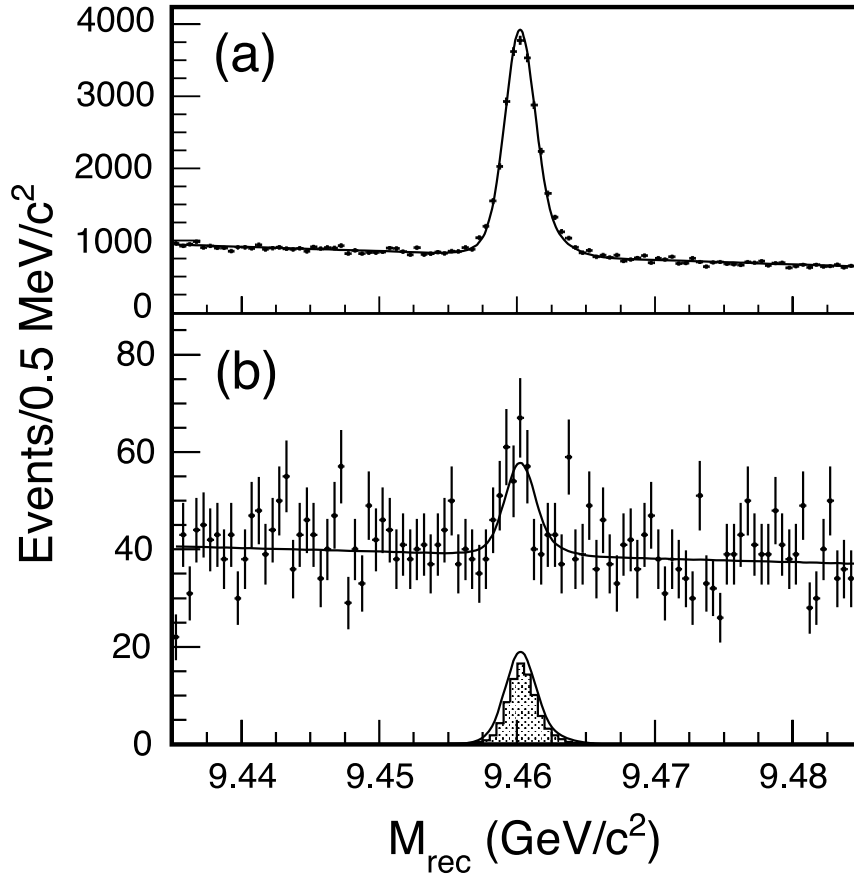}
	\caption{Distribution of the di-pion recoil mass, $M_{\rm rec}$, in the inclusive data sample passing the di-pion selection (a), and in the exclusive data sample, requiring no extra tracks and showers in the event (b). The fit function, a double Gaussian plus a linear background, is superimposed on the data as a solid curve. On the lower plot, the signal shape together with the expected peaking background from $\Upsilon(1S) \to \ell^+\ell^-$ ($\ell = e, \mu, \tau$) decays (shaded histogram) are also displayed.}
	\label{fig:Fig2}
\end{figure}

Figure \ref{fig:Fig2}a shows the distribution of masses recoiling against the di-pion in the inclusive sample. This $M_{\rm rec}$ spectrum is fitted with a double Gaussian signal shape plus a linear background with all parameters allowed to float. The resolution of the core Gaussian is $1.04 \pm 0.04$ MeV/$c^2$ and its fractional area is ($64 \pm 6$)\% of the total. The width of the wider Gaussian, which arises from multiple scattering of charged particles, is twice as much as that of the core Gaussian.  The fit gives a yield of $N[\pi^+\pi^-\Upsilon(1S)] = 18,905 \pm 213 \pm 400$ events, which represents the number  of inclusive $\pi^+\pi^-\Upsilon(1S)$ events. The systematic uncertainty on the number of $\Upsilon(1S)$ events in the inclusive sample is determined from variations of the fit to the recoil mass spectrum. We tried different bin sizes, background parametrization, and signal shape.

The recoil mass distribution in the exclusive sample with no extra tracks and showers larger than 250 MeV is displayed in Fig.~\ref{fig:Fig2}b. The spectrum is fitted with the same signal shape by fixing the parameters of the double Gaussian to be those of the inclusive fit. The yield obtained for invisible $\Upsilon(1S)$ decays is $116 \pm 24 \pm 9$ events. The systematic uncertainty is due to fit variations.

From the study of the generic $\Upsilon(2S)$, continuum, and tau-pair MC samples we conclude that the smooth combinatoric background underneath the signal peak in data with only di-pion selection (Fig.~\ref{fig:Fig2}a) is mainly composed of $\Upsilon(2S)$ resonance decays (75-85\%) and continuum $e^+e^- \to q\bar{q}$ ($q=u,d,s,c$) events (10-15\%). 
Because the background from non-resonant tau-pair events ($e^+e^- \to \tau^+\tau^-$) is negligible, the remaining 5-10\% of the non-peaking background is attributed to two-photon fusion. On the other hand, the exclusive selection on the extra tracks and showers suppresses the combinatoric background from $\Upsilon(2S)$ resonance decays, continuum, and tau-pair events to a negligible level, and therefore the remaining combinatoric background (Fig.~\ref{fig:Fig2}b) is almost exclusively produced by untagged two-photon fusion. 
This is confirmed by the distribution of $\cos\theta_{\pi\pi}$ (without selection on this variable), which peaks along the beam axis as expected for two-photon fusion events.

The MC simulation of $\Upsilon(2S)$ decays also demonstrates that there is an irreducible peaking background in the exclusive sample due to $\Upsilon(1S)$ decays into visible particles when those particles lie outside the detector acceptance. According to the MC simulation, about 93\% of this background is caused by two body decays to $e^+e^-$ and $\mu^+\mu^-$, and 5\% due to decays to $\tau^+\tau^-$.
Other $\Upsilon(1S)$ decays that could fake an invisible signal are estimated to be negligible in comparison to these di-lepton final states.

\begin{figure}
	\centering
	\includegraphics[width=5.5in]{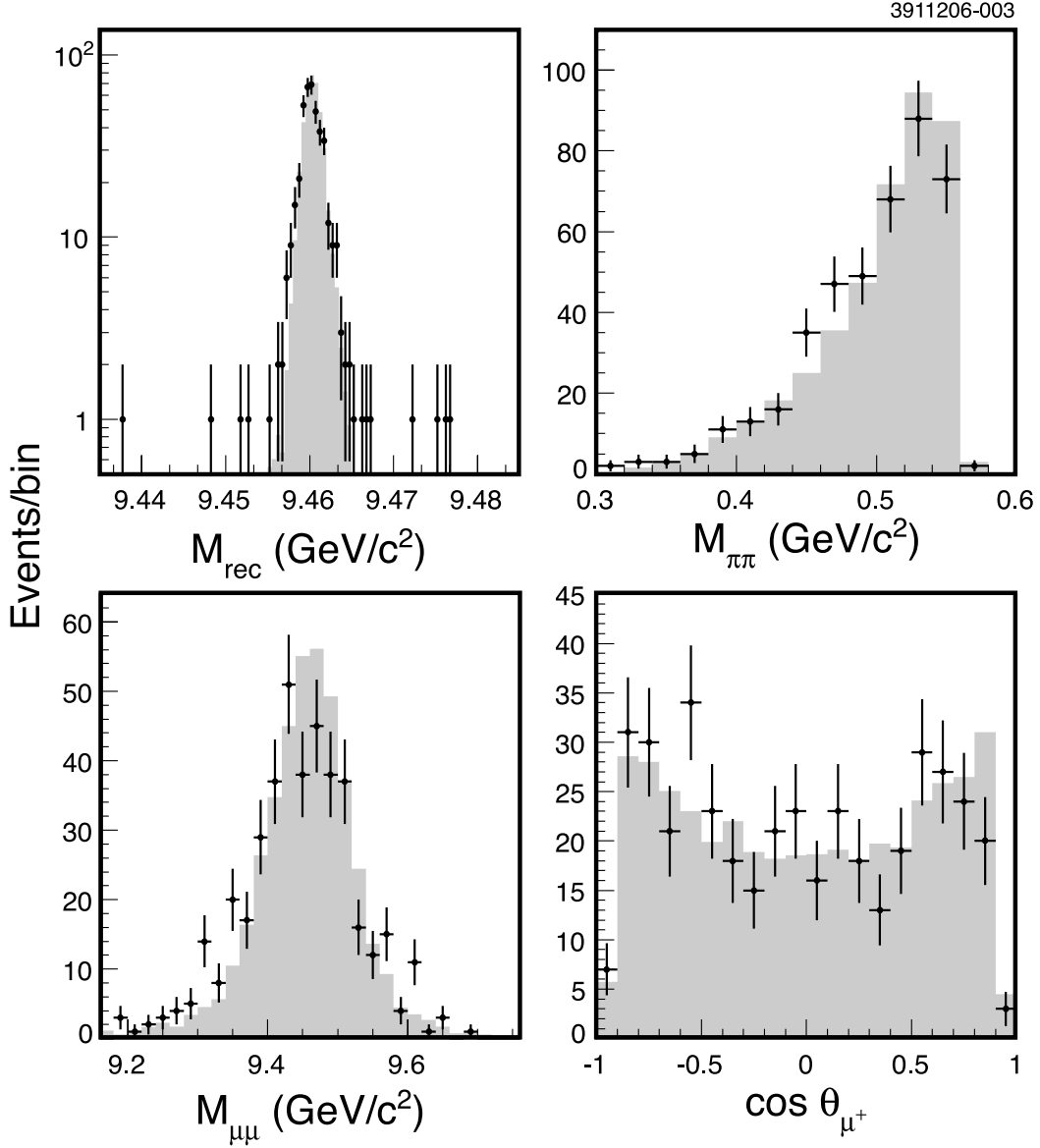}
	\caption{Di-pion recoil mass, di-pion invariant mass, $\mu^+\mu^-$ invariant mass, and the cosine of the polar angle of the $\mu^+$ candidate in fully reconstructed $\Upsilon(2S) \to \pi^+\pi^- \Upsilon(1S)$, $\Upsilon(1S) \to \mu^+\mu^-$ data with two-track trigger requirement (dots) and the corresponding MC simulation (shaded histogram).}
	\label{fig:Fig3}
\end{figure}

In order to get a better estimate of the dominant peaking background due to $\Upsilon(2S) \to \pi^+\pi^- \Upsilon(1S)$, $\Upsilon(1S) \to \ell^+\ell^-$ decays, we use a MC sample generated with EvtGen using the correct angular correlations in decays and simulating final state radiation (the latter of which is particularly important for leptons). The agreement between fully reconstructed $\pi^+\pi^-\mu^+\mu^-$ data and the corresponding MC simulation is demonstrated by Fig.~\ref{fig:Fig3}.

Based on the MC simulation, the acceptance ($\epsilon_{\ell\ell}$) of the exclusive requirement on the extra tracks and showers not associated with the pions for $\pi^+\pi^-\ell^+\ell^-$ events that passed the inclusive di-pion selection is ($8.9 \pm 0.3$)\% with electrons, ($8.5 \pm 0.3$)\% with muons, and ($0.8 \pm 0.1$)\% with tau leptons in the final state. We estimate the number of invisible $\Upsilon(1S) \to \ell^+\ell^-$ events for each lepton flavor ($\ell = e, \mu, \tau$) in the data using the formula
\begin{equation}
\label{eq:invisible_leptonic_decay}
N[\pi^+\pi^-\Upsilon(1S)] \times {\cal B}_{\mu\mu} \times \epsilon_{\ell\ell}~,
\end{equation}
where $N[\pi^+\pi^-\Upsilon(1S)]=18,905$ is the observed number of inclusive $\pi^+\pi^-\Upsilon(1S)$ events in the data, and ${\cal B}_{\mu\mu} = 0.0248$ is the decay branching fraction $\Upsilon(1S) \to \mu^+\mu^-$ \cite{PDG}, and we assume lepton universality (${\cal B}_{\mu\mu} = {\cal B}_{ee} = {\cal B}_{\tau\tau}$). 
The expected backgrounds and statistical uncertainties from $\Upsilon(1S) \to e^+e^-, \mu^+\mu^-,$ and $\tau^+\tau^-$ are $41.7 \pm 1.5, 39.9 \pm 1.5$, and $3.8 \pm 0.4$, respectively.
The total invisible background due to $\Upsilon(1S)$ decaying to lepton pairs is $85 \pm 2 \pm 4$ events, where the systematic error includes uncertainties in the ${\cal B}_{\mu\mu}$ branching fraction, MC efficiency, detector simulation and reconstruction code, and MC statistics.

After subtracting the expected background due to $\Upsilon(1S) \to \ell^+\ell^-$ decays from the invisible signal extracted from the fit to the exclusive data sample, we obtain an invisible signal yield of
\begin{equation}
N[\Upsilon(1S) \to \ {\rm invisible}] = 31 \pm 24 \pm 10.
\label{eqn:Ninvisible}
\end{equation}
The largest systematic uncertainty in ${\cal B}[\Upsilon(1S) \to \ {\rm invisible}]$ comes from the 32\% systematic uncertainty in Eqn.~\ref{eqn:Ninvisible}.
There are additional systematic uncertainties associated with the trigger, the di-pion finding efficiency, the fit to the inclusive data sample, and the additional selection criteria of the exclusive sample.  The combined relative systematic uncertainty in ${\cal B}$ is 34\%.

When we calculate the branching fraction we assume that the trigger efficiency as well as the di-pion finding efficiency is the same in the exclusive and inclusive samples, and thus cancel in the ratio.

There are indications that more tracks and/or showers in the final state can result in the two-track trigger prescale counter incrementing more than once in an event. In particular, we measured the two-track trigger prescale factor to be 18.3 instead of the expected 20 in inclusive $\pi^+\pi^-\Upsilon(1S)$ events by taking the ratio of signal yield from the fit to the di-pion recoil spectrum with and without two-track trigger requirement. 
We cannot directly measure the prescale factor for our signal events with only two soft pions since there is no other trigger requirement that could be always satisfied by these events. Therefore, we assign a 10\% systematic error to the branching fraction due to uncertainty in the ratio of the prescale between the inclusive and exclusive $\pi^+\pi^-\Upsilon(1S)$ events.

We have also measured the AXIAL tracking trigger efficiency for pions with $p_{T} > 150$ MeV/$c$ in fully reconstructed $\Upsilon(2S) \to \pi^+\pi^- \Upsilon(1S)$, $\Upsilon(1S) \to e^+e^-/\mu^+\mu^-$ data where the extra leptons can also satisfy other independent trigger requirements and we do not rely on the two-track trigger for detection. We found an efficiency exceeding $99.9$\%, and therefore, we assume that the systematic uncertainty is negligible due to AXIAL trigger inefficiency.

Based on signal MC simulation, the overall efficiency of the di-pion selection criteria (without the two-track trigger requirement) for invisible $\Upsilon(1S)$ decays is $\sim 27$\%. From comparison to MC simulation of $\Upsilon(2S)$ decays, we assign a 5\% systematic uncertainty due to the relative difference in the di-pion finding and reconstruction efficiency between exclusive $\Upsilon(1S)$ decays and inclusive $\Upsilon(1S)$ decays which have significantly larger track and neutral multiplicity.

The systematic error associated with the selection efficiency of the exclusive cuts on the number of extra tracks and showers with $E>250$ MeV for signal events is estimated by comparing the efficiency for signal MC simulation to that for fully reconstructed $\Upsilon(2S) \to \pi^+\pi^-\Upsilon(1S)$, $\Upsilon(1S) \to e^+e^-/\mu^+\mu^-$ data and the corresponding MC simulation (Fig.~\ref{fig:Fig1}c-d). Based on the differences we assign a conservative systematic error of 1\%.

After correcting the number of invisible signal events with the efficiency of the exclusive selection on the extra tracks and showers in the event ($0.988$), and normalizing with the number of total $\Upsilon(1S)$ decays observed in our sample of $\pi^+\pi^-\Upsilon(1S)$ events ($18,905$), we obtain the branching fraction of invisible $\Upsilon(1S)$ decays as 
\begin{equation}
{\cal B}[\Upsilon(1S) \to \ {\rm invisible}] = (0.16 \pm 0.13 \pm 0.05)\%~ .
\end{equation}
The first error is statistical, from the 77\% relative statistical error in Eqn.~\ref{eqn:Ninvisible},  and the second is systematic.

We then calculate a 90\% confidence level upper limit from the combined (in quadrature) statistical and systematic error using Feldman and Cousins' approach \cite{Feldman} assuming Gaussian distributed error and excluding the non-physical region below zero:
\begin{equation}
{\cal B}[\Upsilon(1S) \to \ {\rm invisible}] < 0.39\% \ (90\% \ {\rm C. L.}).
\end{equation}
Our upper limit is about 50\% larger than the upper limit reported by the Belle experiment \cite{BELLE}. This is consistent with our lower statistics, which is significantly reduced by the trigger prescale on the two-track trigger requirement.

We gratefully acknowledge the effort of the CESR staff in providing us with excellent luminosity and running conditions. D.~Cronin-Hennessy and A.~Ryd thank the A.P.~Sloan Foundation. This work was supported by the National Science Foundation, the U.S. Department of Energy, and the Natural Sciences and Engineering Research Council of Canada.

\end{document}